\documentclass[useAMS,usenatbib,usegraphicx,psfig]{mn2e}
\title[The flaring Nova Scuti 2019]{Multiple flares caused by mass ejection episodes during the advanced nebular phase of Nova Scuti 2019}
\author[Ulisse Munari et al.]
{U. Munari,$^{1}$\thanks{E-mail: ulisse.munari@inaf.it}
G.~L. Righetti,$^{2}$ and
S. Dallaporta,$^{2}$
\\
$^{1}$INAF Astronomical Observatory of Padova, 36012 Asiago (VI), Italy\\
$^{2}$ANS Collaboration, c/o Astronomical Observatory, 36012 Asiago (VI), Italy\\
}

\date{Accepted XXX. Received YYY; in original form ZZZ}

\pubyear{2020}

\begin{document}
\label{firstpage}
\pagerange{\pageref{firstpage}--\pageref{lastpage}}
\maketitle

\begin{abstract}
Our photometric and spectroscopic monitoring shows that starting with 2020
June 4, day +217 from optical maximum and well into its advanced nebular
stage, Nova Sct 2019 begun displaying a series of nine large amplitude
flares (up to $\Delta m$$\sim$1.7 mag), characterized by a rapid rise to
peak ($\leq$10 hours) and a fast exponential decline ($e$-folding
time $\sim$50 hours).  The time interval $\Delta t$ between flares 
follows an ordered sequence, declining from 8.43 to
4.90 days, that safely allows to exclude that any other
flare occured without being recorded by the observations.  When the
sequence of flares was over by 2020 July 28 (day +271), Nova Sct 2019 slowed
its overall decline rate from $\Delta m$=0.0067 mag/day to 0.0027 mag/day. 
The flares were caused by material expelled at high velocity ($\sim$1000
km/s) from the still burning WD.  The cooler pseudo-photosphere forming at
each flare in the expelled material, resulted in a recombination wave to
spread through the original nova ejecta (at $\sim$170 AU from the WD),
quenching emission from [FeX] and [FeVII] and boosting that from lower
ionization species.  After each flare, once the small amount of expelled
material had turned optically thin, the original nova ejecta resumed
displaying [FeX] and [FeVII] emission lines, a fact that clearly proves the
direct photo-ionization action exerted on the ejecta by the burning WD. 
While the other known flaring novae (V458 Vul, V4745 Sgr, and V5588 Sgr)
presented the flares close to maximum brightness and with increasing $\Delta t$,
Nova Sct 2019 is unique in having displayed them during the advanced nebular
stage and with decreasing $\Delta t$.
\end{abstract}
\begin{keywords}
stars: novae, cataclysmic variables
\end{keywords}

\section{Introduction}

The details of the multiple discoveries and designations of Nova Scuti 2019
(NSct19 for short) were given by \citet{green}.  The transient was first
discovered by K.~Nishiyama (Japan) on Oct 29.397 UT (HJD 2458785.897)
at unfiltered 9.4 magnitude, resulting in transient designation
TCP~J18395972-1025415 when reported to CBAT, and independent discoveries by
H.  Nishimura and S.  Kaneko (also from Japan) were soon reported to CBAT by
S.  Nakano.  On Oct 29.524 UT, via VSNET-alert 23669, P.  Schmeer noted the
coincidence of TCP~J18395972-1025415 with the new and unclassified transient
ASASSN-19aad discovered by the ASASSN survey on Oct 29.05 (HJD
2458785.55) at $g$=11.5 mag.  Schmeer also noted the positional
coincidence with a feeble progenitor source recorded by PanSTARR-S1 at
$i$=20.8 and $z$=20.0 mag (and undetected in $g$ and $r$), and based on the
apparent great outburst amplitude, he suggested the object likely to be a
nova.  Confirmation as a nova came soon afterward by
\citet{2019ATel13241....1W} via spectroscopic observations obtained on Oct
29.81 UT with the Liverpool telescope, which revealed several broad emission
lines flanked by P-Cyg absorptions, features that were also noted by
\citet{2019ATel13245....1P} on their spectroscopic observations for Oct
29.54 UT.  A description of the early evolution of profiles of emission
lines and associated P-Cyg absorptions was provided by
\citet{2020AN....341..781J}, who recorded a series of six high resolution
spectra covering the first $\sim$two weeks of the outburst.  The TNS server
assigned the designation AT\,2019tpb to the transient and the General
Catalog of Variable Stars (GCVS) provided the permanent designation
V659~Sct.

    \begin{figure*}
	\includegraphics[angle=270,width=17.6cm]{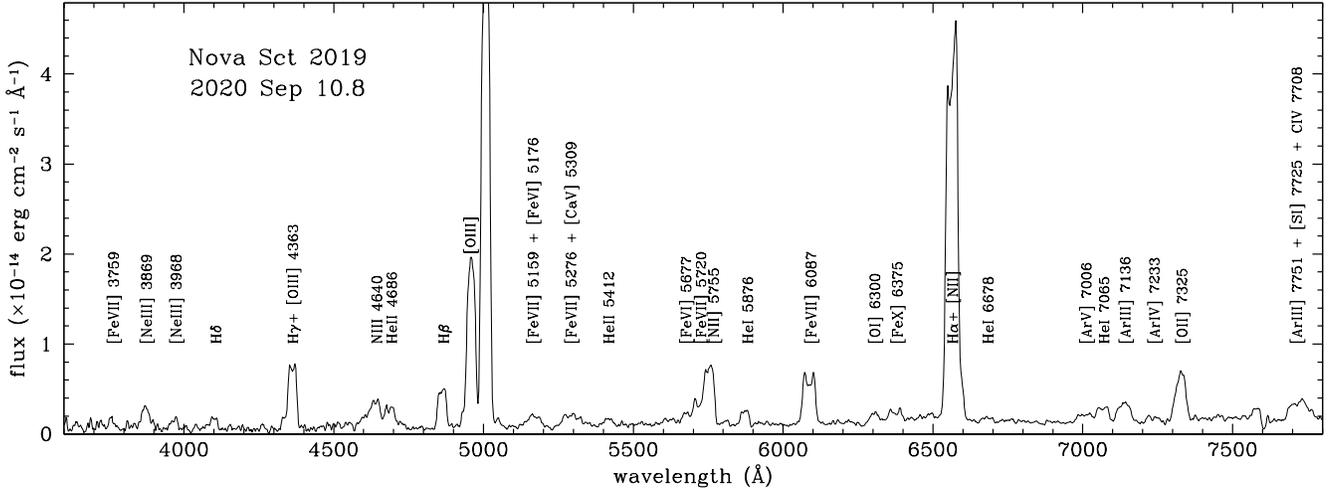}
        \caption{Spectrum for 10 Sept 2020 (day +316 from outburst maximum),
        representative of all the others we have recorded during our 2020
        observing campaign of Nova Sct 2019, while it was going through its
        advanced nebular decline.  The strongest lines are identified.}
    \label{fig:fig1}
    \end{figure*}

Not much else was reported about NSct19 during the rest of 2019, with the
exception of a pointing with the X--ray/UV {\it Swift} satellite by
\citet{2019ATel13252....1S}, that did not detect the nova in the X--rays and
recorded it at UVM2=14.27 mag in the ultraviolet, which led them to estimate
the reddening as $E_{B-V}$$\sim$0.9.

The interest in NSct19 was briefly renewed the following year by
\citet{2020ATel13815....1W} that reported detecting [SiVI], [SiVII],
[CaVII] and [FeIX] coronal emission lines in an infrared spectrum of the
nova they recorded on 2020 June 6 with IRTF.  This triggered our interest in
the nova and by the following day, 2020 June 19, we begun a
$B$$V$$R$$I$ photometric and 3300-8000~\AA\ spectroscopic monitoring that
covered the rest of the observing season for the nova up to its Solar
conjunction in late October 2020.  \citet{2020ATel13819....1S} obtain a
low-resolution optical spectrum of NSct19 on 2020 June 19, revealing the
nova to be well into its nebular stage, and confirmed the high excitation
conditions seen by \citet{2020ATel13815....1W} in the infrared by detecting
the coronal [FeX] 6375 \AA\ in emission among a rich assortment of
double-peaked emission lines distributed over a wide range of ionization
conditions (from [OI] to [FeVII]).

In this paper we discuss the results of our monitoring of NSct19 during
2020, which coincide with its advanced nebular decline, focussing in
particular on the surprising appearance of a series of very fast ($\sim$2
days duration) and large amplitude flares (up to $\Delta B$=1.7 mag).

        \begin{table}
	\centering
                \caption{Our BVRI photometry on the Landolt system of Nova
                Sct 2019.  The complete table is only available in
                electronic as supplementary material; a portion is
                shown here to provide guidance on its content.  HJD is the
                heliocentric JD$-$2450000.}
                \label{tab:tab1}
        \small
    \begin{tabular}{@{}c@{~}c@{~}c@{~}c@{~}c@{~}c@{~}c@{~}c@{~}c@{~}c@{}}
    \hline
	&&\\
      HJD   &   B   &   err &   V   &   err &   R    & err  &   I   &  err   &  ID   \\
	&&\\
 9023.483 &15.799 & 0.024 &14.778 & 0.023 &14.038  &0.014 &13.887 & 0.015  & 1507  \\ 
 9023.491 &15.769 & 0.012 &14.729 & 0.013 &14.033  &0.010 &13.801 & 0.012  & 0310  \\ 
 9024.408 &15.884 & 0.017 &14.908 & 0.015 &14.285  &0.012 &14.133 & 0.024  & 1507  \\ 
 9024.496 &15.924 & 0.014 &14.931 & 0.013 &14.279  &0.010 &14.173 & 0.012  & 0310  \\ 
	&&\\
    \hline
    \end{tabular}
        \end{table}

    \begin{table}
    \centering
    \caption{Log spectroscopic observations recorded with Asiago 1.22m + B\&C + 300
    ln/mm (3300-8000 \AA, 2.3 \AA~pix$^{-1}$).}
    \label{tab:tab2}
    \begin{tabular}{c@{~~}c@{~~}ccrc}
    \hline
      &&\\
      \multicolumn{3}{c}{date}& UT & expt & HJD \\
      &&& middle & (sec) & (-2450000) \\
      &&\\
	2020 & Jun & 28 & 22:34 & 3600 & 9029.440 \\
	2020 & Jul & 05 & 00:03 & 4800 & 9035.502 \\
	2020 & Jul & 06 & 00:02 & 5400 & 9036.502 \\
	2020 & Aug & 20 & 21:37 & 1080 & 9082.401 \\
	2020 & Aug & 25 & 20:49 & 1080 & 9087.367 \\
	2020 & Aug & 27 & 19:39 & 1140 & 9089.319 \\
	2020 & Sep & 09 & 19:48 & 1200 & 9102.325 \\
	2020 & Sep & 10 & 20:03 & 1800 & 9103.335 \\
	2020 & Sep & 11 & 19:52 & 1200 & 9104.328 \\
      &&\\
    \hline
    \end{tabular}
    \end{table}

\section{Observations}

We have obtained $B$$V$$R$$I$ optical photometry of NSct19 in the
\citet{2009AJ....137.4186L} photometric system from 2020 June 19 to
September 18 at $\sim$daily cadence, and then three more every $\sim$ten
days to October 16, with ANS Collaboration telescopes ID 0310 and 1507;
when close in time their data were not combined, to provide a mutual
check. Data reduction has involved all the usual steps for bias, dark and
flat with calibration images collected during the same observing nights.  We
adopted aperture photometry because the sparse field around NSct19 did not
required a PSF-fitting approach.  The transformation from the local to the
Landolt standard system was carried out via nightly colour equations
calibrated on a photometric sequence recorded on the same frames with NSct19
and extracted from the APASS DR8 survey \citep{2014CoSka..43..518H}, ported
to the \citet{2009AJ....137.4186L} system via the transformations calibrated
by \citet{2014AJ....148...81M}.  Our photometry of NSct19 is listed in
Table~\ref{tab:tab1}.  The quoted errors are the quadratic sum of the
Poissonian error on the variable and the error in the transformation to the
standard system via colour equations.

Low resolution spectroscopy of NSct19 has been obtained with the 1.22m
telescope + B\&C spectrograph operated in Asiago by the Department of
Physics and Astronomy of the University of Padova.  The CCD camera is a
ANDOR iDus DU440A with a back-illuminated E2V 42-10 sensor, 2048$\times$512
array of 13.5 $\mu$m pixels.  A 300 ln/mm grating blazed at 5000~\AA\
results in 2.3~\AA~pix$^{-1}$ dispersion and 3300-8000~\AA\ spectral
coverage.  The slit has always been rotated to the parallactic angle for
optimal flux mapping.  All data have been similarly reduced within IRAF,
carefully involving all steps connected with correction for bias, dark and
flat, sky subtraction, wavelength calibration and heliocentric correction. 
The spectra have been flux calibrated against observations of the nearby
spectrophotometric standard HR~7032 (2$^\circ$ angular distance) observed
each night immediately before or after NSct19, and the zero-points checked
against the result of simultaneous ANS Collaboration $B$$V$$R$$I$
photometry.  A log of the spectroscopic observations is given in
Table~\ref{tab:tab2}.

        \begin{table}
	\centering
                \caption{Summary of basic parameters for Nova Sct 2019.} 
                \label{tab:tab3}
        \small
    \begin{tabular}{ll}
    \hline
	&\\
names              & Nova Sct 2019        \\
                   & V659 Sct             \\       
                   & TCP~J18395972-1025415\\
                   & ASASSN 19aad         \\
                   & AT 2019tpb           \\
equatorial         &RA = 18:39:59.82      \\
                   &DEC = $-$10:25:41.9   \\
Galactic           &$l$ =   022.352       \\
                   &$b$ = $-$02.227       \\
outburst:   maximum &UT = 2019 Oct 31.0    \\
                   &$V$ = 8.38 mag        \\
                   &type = FeII           \\
\multicolumn{1}{r}{decline} &$t_2$ = 7.0 days      \\
                   & $t_3$ = 13.5         \\
\multicolumn{1}{r}{amplitude} &$\Delta I$$\sim$14.4 mag\\ 
         reddening &$E_{B-V}$ = 1.1       \\
distance (from $t_3$)& 5.3 kpc            \\
	&\\
    \hline
    \end{tabular}
        \end{table}

    \begin{table}
    \centering
    \caption{Average values for the interstellar KI line we measured on 
    Jack et al.  (2020) spectra of Nova Sct 2019.}
    \label{tab:tab4}
    \begin{tabular}{c@{~}ccc@{~}cc}
    \hline
      &&\\
      \multicolumn{2}{c}{RV$_\odot$} &  FWHM  & \multicolumn{2}{c}{equiv. width} & $E_{B-V}$ \\  
      \multicolumn{2}{c}{(km/s)}     & (km/s) & \multicolumn{2}{c}{(\AA)}        & (mag) \\  
      &&\\
      $-$8.3 & $\pm$0.2  & 19 & 0.206 & $\pm$0.003 & 0.81 \\
      $+$33.3& $\pm$1.2  & 29 & 0.080 & $\pm$0.004 & 0.30 \\
      &&\\
    \hline
    \end{tabular}
    \end{table}

    \begin{table}
    \centering
    \caption{FWHM (corrected for instrumental resolution) and velocity separation 
    of double peaks for some representative lines in the spectrum of
    Figure~\ref{fig:fig1}.}
    \label{tab:tab5}
    \begin{tabular}{lcc}
    \hline
      &&\\
      line & FWHM & peaks\\
           & (km s$^{-1}$) &(km s$^{-1}$) \\
      &&\\
      HeI 7065    & 1930 & 1090 \\
      H$\alpha$   & 1950 & 1280 \\
      HeII 4686   & 2200 & 1320 \\
      $[$OIII$]$ 4363 & 2200 & 1380 \\
      $[$FeVII$]$ 6987& 2210 & 1450 \\
      $[$FeX$]$ 6375  & 2450 & 1550 \\
      &&\\
      $[$OI$]$ 6300   & 1430 &      \\
      $[$OII$]$ 7325  & 1480 &      \\
      $[$OIII$]$ 5007 & 1930 &      \\
      $[$NeIII$]$ 3869& 2550 &      \\
      &&\\
    \hline
    \end{tabular}
    \end{table}

\section{Basic parameters of Nova Sct 2019}

Observing conditions at the time of discovery were far from ideal, with the
object low on the horizon for the fast approaching Solar conjunction, and the
Moon at short angular distance.  Interpolating the AAVSO lightcurve with a
spline function, the time of maximum in $V$-band is derived as Oct 31.0
($\pm$0.5) UT at $V$=8.38($\pm$0.1), with $B$ band anticipating by
$\sim$half a day and $R$, $I$ bands delayed by $\sim$half a day from
$V$-band, as expected from an expanding fireball
\citep[eg.][]{2017MNRAS.469.4341M}.  A spectrum taken shortly after maximum
brightness is available in the ARAS database \citep{2019CoSka..49..217T},
and it shows NSct19 belonging to the FeII-class of novae
\citep{1992AJ....104..725W}.  ASASSN observed the field of NSct19 for only a
couple of days into the outburst because of the approaching Solar
conjunction, deriving $g$$\geq$17.0, $g$=11.57, and $g$=9.54 for Oct 28.06,
Oct 29.06, and Oct 30.06 UT, respectively.

\citet{1987A&AS...70..125V} list +0.23 and $-$0.02 as the intrinsic
(B-V)$_\circ$ colour for novae at maximum and $t_{2}^{V}$, respectively. 
From a spline interpolation of the AAVSO lightcurve we estimate
$B$$-$$V$$\sim$1.56($\pm$0.15) at maximum and $B$$-$$V$$\sim$1.1($\pm$0.10)
at $t_{2}^{V}$, corresponding to $E_{B-V}$=1.3 and 1.1 mag, respectively.  VSNET
CCD photometry for Oct 30.07 UT (observer K.  Yoshimoto) provides $V$=8.59 and
$B$$-$$V$=1.20, resulting in $E_{B-V}$=1.0 mag.  \citet{2020AN....341..781J}
reported about saturated multi-components for the interstellar NaI lines
recorded on their TIGRE high resolution spectra.  On our request, D.  Jack
kindly forwarded us such spectra, and on them we measured the unsaturated,
multi-component profile of the KI 7699~\AA\ interstellar line, obtaining the
values listed in Table~\ref{tab:tab4}.  By adopting the calibration of
\citet{1997A&A...318..269M}, the equivalent width of the two KI components
translate into a total reddening $E_{B-V}$= 0.81 + 0.30 = 1.11 mag.  On the
same TIGRE spectra, we also measured an average 0.2568~\AA\ equivalent width
for the diffuse interstellar band (DIB) at 6614 \AA.  Adopting for such DIB
the calibration by \citet{2014ASPC..490..183M}, its equivalent width
translates to $E_{B-V}$=1.13 mag.  Averaging over the various estimates, we
derive $E_{B-V}$=1.10 ($\pm$ 0.05) as the reddening affecting NSct19.

\citet{2019A&A...622A.186S} have re-calibrated the standard MMRD relation
(mag at maximum vs.  rate of decline) on GAIA DR2 parallaxes.  Applying it
to the above $t_{3}^{V}$=13.5($\pm$0.7) days leads to an absolute magnitude
M(V)=$-$8.7 and, by combining with $E_{B-V}$=1.10 for a standard $R_V$=3.1
reddening law \citep{1999PASP..111...63F}, to a distance of 5.3 kpc to
NSct19.  At such a distance, the Bayestar2019 3D model of Galactic
extinction by \citet{2019ApJ...887...93G} returns $E_{g-r}$$\geq$0.96.

Table~\ref{tab:tab3} summarizes the basic parameters of NSct19.

\section{Bright flares during the advanced nebular phase}

    \begin{figure}
	\includegraphics[width=8.7cm]{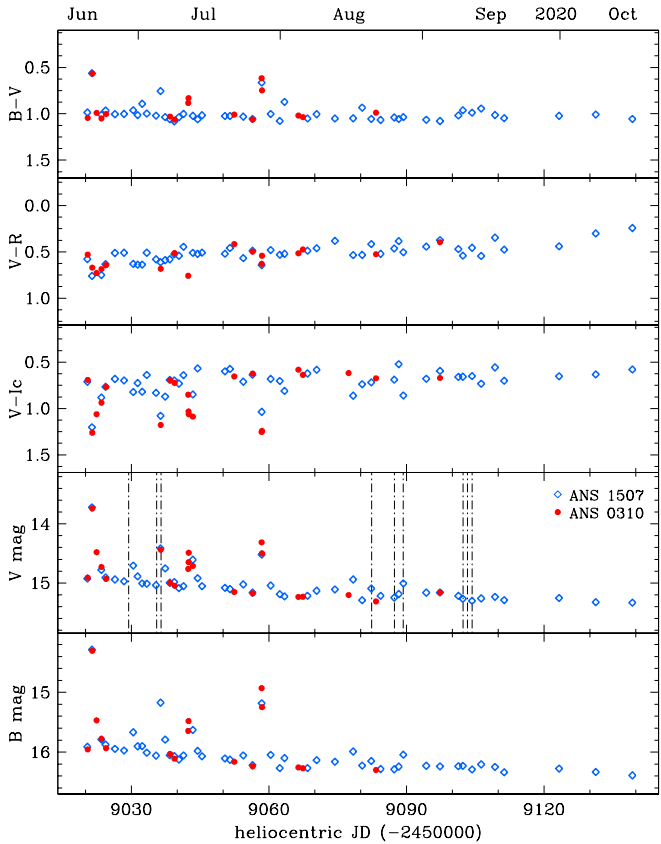}
        \caption{Photometric evolution of Nova Sct 2019 during the advanced
        nebular phase from our $B$$V$$R$$I$ photometric observations in
        June-October 2020 (Table~\ref{tab:tab1}).  The vertical lines in the $V$-band panel mark
        the epochs of our spectra listed in Table~\ref{tab:tab2}.}
    \label{fig:fig2}
    \end{figure}

When we began our monitoring of NSct19 in June 2020, the nova was already
well into its advanced nebular stage as illustrated by the spectrum in
Figure~\ref{fig:fig1}, where [OIII] 4959, 5007 stand out prominently and
stronger than H$\alpha$.  Considering the large intensity of [NII] 5755,
some unresolved emission from [NII] 6548, 6584 probably contribute to the
H$\alpha$ profile.  The spectrum in Figure~\ref{fig:fig1} is well
representative of our entire June-October observing period, and well
supports the reports by \citet{2020ATel13815....1W} and
\citet{2020ATel13819....1S} for the presence of coronal emission lines in
the infrared and optical spectra of the nova they recorded in June 2020. 
[FeX]~6375 is clearly present in our spectra (for a recent census of novae
showing [FeX] see \citet{2021AJ....161..291R}), as well as a full
assortments of [FeVI] and [FeVII] lines in addition to [ArIII], [ArIV] and
[ArV] transitions.  Their profiles range in shape from Gaussian-like to
double-peaked, with Table~\ref{tab:tab5} listing the FWHM (corrected for
instrumental resolution) and separation of peaks for some representative
lines.  Our FWHMs are remarkably close to those measured by
\citet{2020AN....341..781J} during the first two weeks of the outburst.  No
emission component is visible in our spectra to match the higher velocities
characterizing the P-Cyg absorptions tracked by \citet{2020AN....341..781J}.

    \begin{figure*}
	\includegraphics[width=17.6cm]{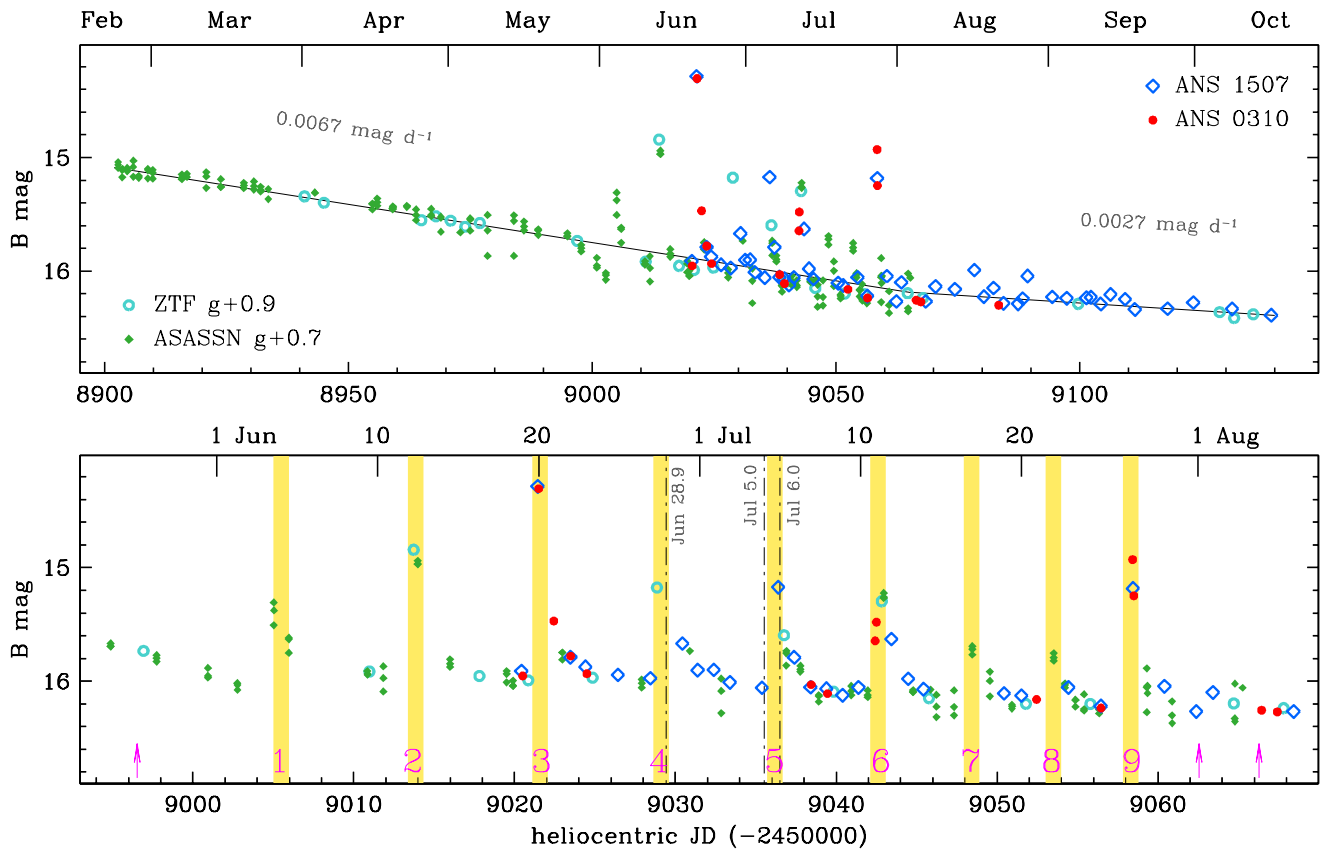}
        \caption{{\it Top panel}: $B$-band lightcurve of Nova Sct 2019 
        covering 2020, built from our data in Table~\ref{tab:tab1} and
        and $g$-band data from ASASSN and ZTF sky-patrol surveys, scaled
        by the indicated quantities to match the $B$-band data. {\it
        Bottom panel}: zooming onto the June-July portion of the lightcurve
        from the above panel to highlight the 9 flares experienced by the
        nova. The dot-dashed vertical lines mark the epoch of the three spectra
        discussed in Figure~\ref{fig:fig6}. For the meaning of the three pink 
        arrows see text (sect. 4).}
    \label{fig:fig3}
    \end{figure*}

The photometric evolution of NSct19 as recorded by our June-October 2020
observations is presented in Figure~\ref{fig:fig2}.  It is characterized by
a rather slow decline in brightness ($\Delta V$=0.5~mag in 120 days), as
typical of novae during the advanced nebular stage.  

The most striking feature of the NSct19 lightcurve in Figure~\ref{fig:fig2}
is however the presence of a number of {\it flares}, short-lived
brightenings of between 0.4 and 1.7 magnitudes in $V$ and $B$, which are
superimposed onto the otherwise normal and smooth decline.  Such events are
extremely rare in novae (see discussion below in sect.  6.2) and flag NSct19
with special interest.  These flares should not be confused with the jitters
many slow novae present much earlier in their evolution, during the long
plateau they experience around maximum brightness
\citep[eg.][]{2010AJ....140...34S}, or around the transition from optically
thick to thin conditions \citep[eg.][]{1964gano.book.....P}.

To put the detection of flares into context, in Figure~\ref{fig:fig3} we
have built a more comprehensive lightcurve of NSct19 for 2020, by combining
our $B$-band data with $g$-band measurements collected by ASASSN
\citep{2014ApJ...788...48S, 2017PASP..129j4502K} and ZTF patrol surveys
\citep{2019PASP..131a8003M, 2019PASP..131a8002B}, that we have retrieved
from their respective databases.  An offset has been applied to $g$-band
data (+0.7 mag for ASASSN, +0.9 mag for ZTF) to bring them on the same scale
of $B$-band data.

The photometric decline of NSct19 during 2020 has been characterized by two
different slopes, as clearly visible in Figure~\ref{fig:fig3}: an initial
(February through July) faster decline at 0.0067 mag~day$^{-1}$, followed by
a slower one at 0.0027 mag~day$^{-1}$.  The flares appeared at the end
of the faster-decline portion of the lightcurve, and they ceased as NSct19
settled onto the slower-decline descent.  The change in the decline speed
has been probably governed by some adjustment in the recombination vs
photo-ionization of the ejecta, which may have been driven by changes in
rate of nuclear burning on the central WD and/or changes of its out-flowing
wind, and by the possible injection of new material in the inner
circumstellar space as a result of the repeated flares.

\begin{figure}
	\begin{center}
	\includegraphics[width=5.5cm]{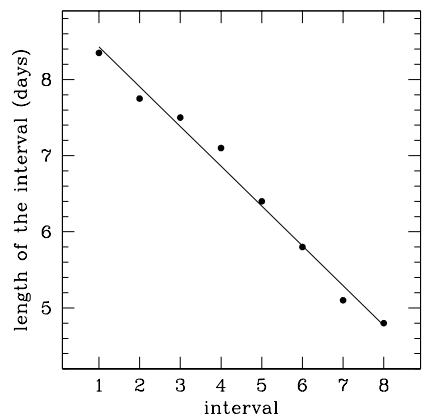}
        \caption{Length (in days) of the eight intervals between the
        successive nine flares listed in Table~\ref{tab:tab6}.}
	\end{center}
    \label{fig:fig4}
    \end{figure}

The lower panel of Figure~\ref{fig:fig3} zooms on the time interval covered
by the flares, which are highlighted by the yellow vertical bands.  They are
9 in number and their epochs are listed in Table~\ref{tab:tab6}.  The flares seem to
follow a precise temporal sequence, as illustrated by Figure~\ref{fig:fig4}
where we have plotted the time interval between two successive flares.  A 
tight linear trend is obvious, and the small deviations from it could be
easily accounted for by the limited accuracy to which the time of maxima can
be derived with the available data (sampling time $\sim$0.5 day). 
Extrapolating the linear trend outside the recorded 9 flares allow to
predict the times of occurrence for any further such event preceding or
following those actually observed, and such times are marked with pink
arrows in the lower panel of Figure~\ref{fig:fig3}.  The arrows coincide in
time with observations that caught NSct19 at the normal quiescence level,
excluding that any further flare has occoured without being recorded by
observations (at least any other flare obeying to the linear trend of
Figure~\ref{fig:fig4}).

    \begin{table}
    \centering
    \caption{Epochs of the nine flares exhibited by Nova Sct 2019.
    They correspond to the brightest photometric observation
    recorded for the given event, by either us, ASASSN, or ZTF. Given the
    sparse sampling, the actual epoch of true peak brightness my differ up
    to $\sim$0.5 day from the listed values.}
    \label{tab:tab6}
    \begin{tabular}{ccc}
    \hline
      &&\\
      flare & HJD        & UT date \\
       N.   & (-2450000) & (2020)  \\
      &&\\
	1  & 9005.302  &  June 4.802 \\ 
	2  & 9013.731  &  June 13.231 \\ 
	3  & 9021.440  &  June 20.940 \\ 
	4  & 9028.855  &  June 28.355 \\ 
	5  & 9036.398  &  July 05.898 \\ 
	6  & 9042.828  &  July 12.328 \\ 
	7  & 9048.455  &  July 17.955 \\ 
	8  & 9053.527  &  July 23.027 \\ 
	9  & 9058.422  &  July 27.922 \\ 
        &&\\
    \hline
    \end{tabular}
    \end{table}

If the $e-$folding time for the brightness decline appears to be similar for
all flares, approximately 50~hours, their amplitude may have not been always
the same, with flare N.3 peaking at $\Delta B$=1.7 and N.7 and 8 limited to
$\Delta B$=0.4 mag.  Any definitive conclusion about flare amplitude is
hampered by the sampling time of the observations ($\sim$0.5 days) and the
real possibility that the true maximum has been missed altogether for some
of them.

The takeaways from this section are: (1) a series of 9 fast-evolving and
large-amplitude flares have been recorded during the advance nebular
decline, just prior to a major change in the rate of decline of NSct19; (2)
the flares are arranged in a time sequence that allows safely to conclude
that all events have been recorded and none has gone unnoticed; and (3) not all
flares attained the same brightness amplitude.

    \begin{figure}
	\includegraphics[width=8.7cm]{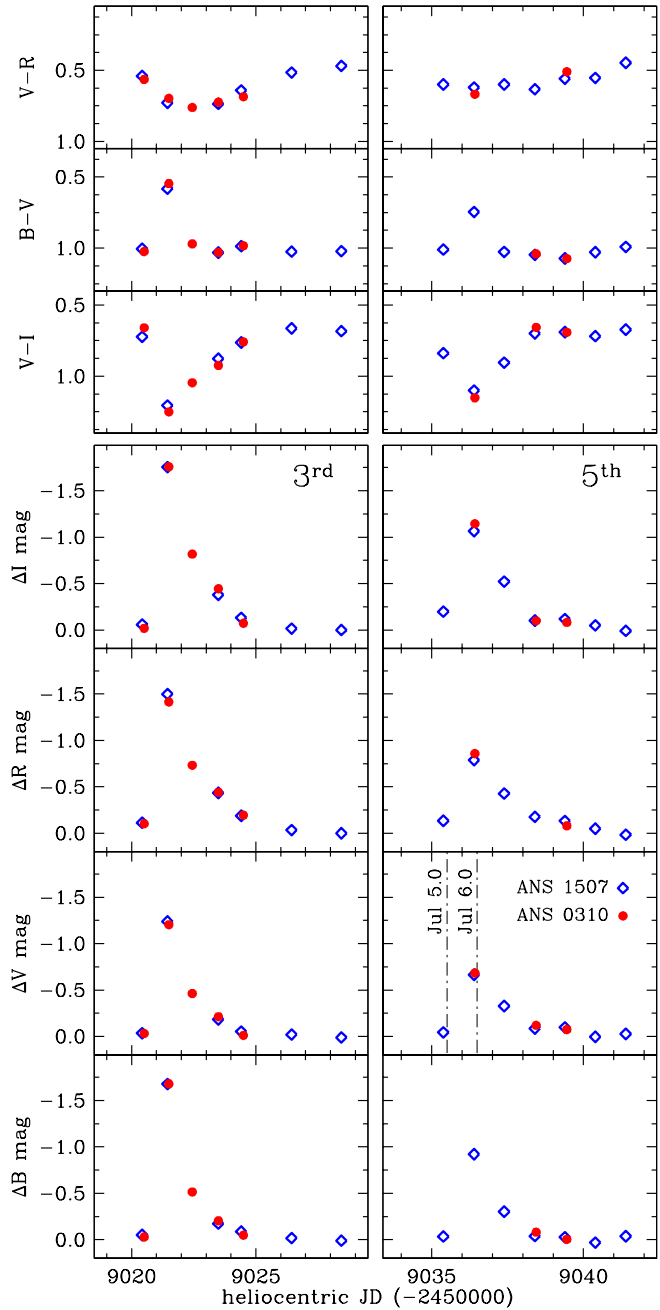}
        \caption{Expanded view of the lightcurve in Figure~\ref{fig:fig2}
        to highlight the evolution of Nova Sct 2019 during the 3rd and 5th
        flares, those best covered by our $B$$V$$R$$I$ observations.}
    \label{fig:fig5}
    \end{figure}

\section{Anatomy of the flares}

The observations we collected allow to document in detail the photometric
and spectroscopic characteristics of the flares experienced by NSct19.  The
flares best covered by our photometry are the 3rd and the 5th, and
Figure~\ref{fig:fig5} provides a zoom on their $B$$V$$R$$I$ light- and
colour-curves: the two flares behaved very similarly and the apparent 
difference in peak brightness is probably an effect of the sampling by the 
observations.

\subsection{Photometric properties}

The rise to maximum brightness for all flares has been very fast.  Our
sampling time interval put a general $\la$1-day upper limit to that.  A more
stringent value can be derived from the evolution of flare N.4 (cf. 
Figure~\ref{fig:fig3}).  When we observed NSct19 on June 27.940 UT it was
still at the normal quiescent level observed between flares, but shortly
afterward, on June 28.355 UT, ZTF caught the nova one magnitude brighter
and close to peak flare brightness, implying an upper limit to the rise to
maximum brightness of $\leq$10~hours.

The photometric colour-change associated to the flares is rather peculiar, as
illustrated by Figure~\ref{fig:fig5}: at peak of the flare, NSct19 becomes
{\it bluer} in $B$$-$$V$ by 0.5 mag, and {\it redder} by 0.5 mag in
$V$$-$$I$, while $V$$-$$R$ remains rather flat.  In other words, the amount
of brightening is larger at both ends of the optical range ($B$ and $I$
bands) than it is in the middle ($V$ and $R$ bands).

\begin{figure*}
	\includegraphics[angle=270,width=17.6cm]{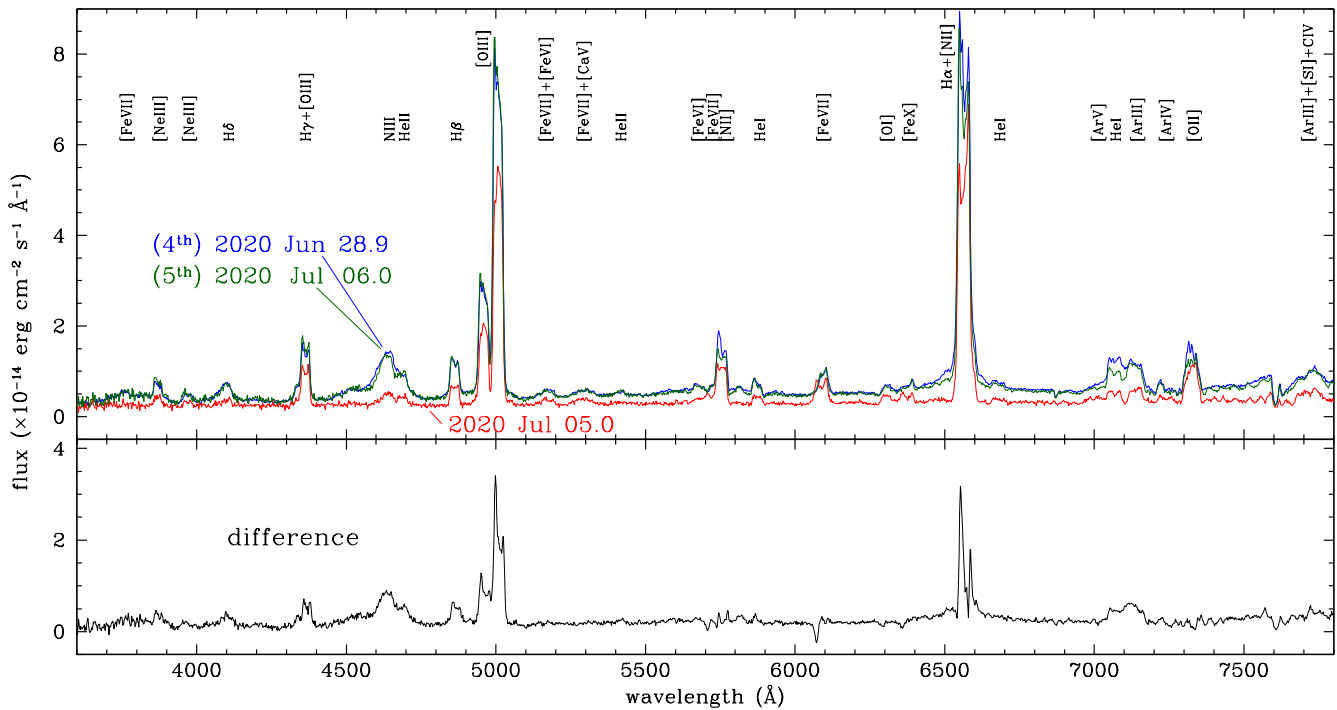}
        \caption{{\it Upper panel:} our spectra obtained at peak of flares
        N. 4 and 5 (June 28.9 and July 6.0 UT), and during the in-between
        quiescence (July 5.0 UT). {\it Lower panel:} result of subtracting the
        quiescence spectrum from the average of the two flare spectra.}
    \label{fig:fig6}
    \end{figure*}

Both the continuum and the emission lines contribute to the flux recorded
through the photometric bands, and disentangling the respective roles played
during flares is impossible on purely photometric grounds.  Accurately
fluxed spectra are required to this aim.  Luckily, two of our spectra (June
28.94 and July 06.00 UT, cf.  Table~\ref{tab:tab2}) were obtained close in
time to the peak of the 4th and 5th flares, while a third (Jul 05.00 UT) was
observed during the brief quiescence in between them (their epochs are
marked by the dot-dashed vertical lines in Figure~\ref{fig:fig3}).  These
three spectra are compared in Figure~\ref{fig:fig6}.  The spectra at flare
maxima look almost identical, as are the colour- and
light-curves for the 3rd and 5th flares presented in Figure~\ref{fig:fig5}. 
Such similarities support the notion that the mechanism driving the flares
has been one and the same throughout the whole series of nine recorded
events.

Given their similarities, we have averaged the two spectra at flare maximum
and subtracted from them the spectrum for the in-between quiescence.  The
resulting difference-spectrum is plotted in the lower panel of
Figure~\ref{fig:fig6}.  We have then integrated the flux of the spectra in
Figure~\ref{fig:fig6} through the transmission profile of the $B$$V$$R$$I$
photometric bands as tabulated by Landolt (1992), with the zero-points being
fixed by repeating an identical operation on the spectrophotometric
standards observed along with NSct19.  The resulting magnitudes are listed
in Table~\ref{tab:tab7}, where we also report the photometric magnitude
corresponding to the flux radiated separately in the continuum (fitted with
a bremsstrahlung distribution) and in the emission lines.  From
Table~\ref{tab:tab7} it is evident how the variation going from quiescence
to flare peak is larger for the continuum (1.0 mag) that it is for the
emission lines (0.4 mag).  A large change affects the {\it 4640-blend},
probably composed of NIII lines pumped by fluorescence from HeII Ly$\alpha$
via OIII 374.432 \AA\ \citep{1947PASP...59..196B, 2007A&A...464..715S}.  The
3$\times$ increase in the intensity of the 4640 blend, which is located
close to the peak of the $B$-band transmission profile, seems the prime
responsible for the larger amplitude (0.8 mag) in $B$ compared to $V$ and
$R$ (both 0.4 mag) for the variation due to the emission lines.

As well illustrated by Figure~\ref{fig:fig6} and Table~\ref{tab:tab7}, in
going from quiescence to flare-peak the underlying continuum brightens
similarly at all (optical) wavelengths, so its shape remained unchanged.  If
the fluorescence-pumped NIII 4640 complex may contribute to a larger
brightening of NSct19 in $B$ band compared to $V$ and $R$, a similar role
could be played for the $I$ band by OI 8446 line, fluorescence-pumped by HI
Ly$\beta$ (Bowen 1947), and located close to the peak of the $I$ band
transmission profile.  The OI 8446 line is rather strong in novae, frequently second
only to H$\alpha$ in terms of emitted flux \citep[eg.][]{2014MNRAS.440.3402M}. 
Unfortunately, we cannot test this hypothesis with our spectra that do not
extend redward of 8000~\AA.

\subsection{Changes in the emission line profiles}

    \begin{table}
    \centering
    \caption{Photometry from flux integration on spectra of Nova Sct 2019
    taken at flare peak (Jun 28.9 and July 6.0) and in-between quiescence
    (Jul 5.0). The central column refers to the spectra zipped of their emission lines,
    and the right column to the spectra subtracted of their underlying continuum 
    (approximated by fitting to a bremsstrahlung distribution).}
    \label{tab:tab7}
    \begin{tabular}{@{}c@{~}c@{~}c@{~}c@{~~}c@{~}c@{~}c@{~}c@{~~}c@{~}c@{~}c@{}}
    \hline
      &&\\ 
    \multicolumn{3}{c}{spectrum}&&\multicolumn{3}{c}{continuum}&&\multicolumn{3}{c}{em. lines}\\ \cline{1-3} \cline{5-7} \cline{9-11}
	           $B$ & $V$ & $R$ && $B$ & $V$ & $R$ && $B$ & $V$ & $R$\\
      &&\\
    \multicolumn{11}{c}{\it flare peak}\\
	          &       &&       &       &       &&       &       &       \\
            15.17 & 14.47 & 13.74 && 16.05 & 15.13 & 14.39 && 15.84 & 15.33 & 14.64 \\
                  &       &&       &       &       &&       &       &       \\
	 \multicolumn{11}{c}{\it quiescence}\\
      &&\\
            16.06 & 15.18 & 14.46 && 16.97 & 16.20 & 15.39 && 16.68 & 15.72 & 15.06 \\
      &&\\
	 \multicolumn{11}{c}{\it difference spectrum}\\
      &&\\
            15.81 & 15.31 & 14.67 && 16.70 & 15.73 & 15.04 && 16.44 & 16.56 & 16.03 \\
      &&\\
    \hline
    \end{tabular}
    \end{table}

The type of change in the profiles of emission lines of NSct19 in going from
quiescence to flare peak is illustrated in Figures~\ref{fig:fig7} and
~\ref{fig:figr}, which zooms on spectra presented in Figure~\ref{fig:fig6}.

For all the emission lines, with the except of [FeVII] and [FeX], the flare
causes the emission profile to develop a blue-peaked component over an
otherwise flat (eg.  [NII] 5755, [ArIII] 7136, HeI 5876) or rounded
top ([OIII] 4959 and 5007, [OII] 7325).  The radial velocity of such blue
peak is $-$650 km/s for all lines, as indicated by the vertical lines in
Figure~\ref{fig:fig7}.

An opposite behavior characterized the highest ionization emission lines:
as illustrated by Figure~\ref{fig:fig7} for [FeVII] 5720,6087 and [FeX]
6375, their line profiles in quiescence are double-peaked, with the blue
peak at $-$650 km/s that disappears during a flare.

Going from quiescence to flare peak also causes the appearance of broad
wings, albeit of low intensity, to permitted lines which have no
counterparts for nebular ones, as deducible from Figure~\ref{fig:fig6} by
comparing H$\alpha$ and [OIII] lines. Figure~\ref{fig:figr} zooms on the
H$\alpha$ wings from Figure~\ref{fig:fig6}, fitted with a simple Gaussian
of FWHM$\sim$2300 km/s.

\section{Discussion}

\subsection{Interpreting the flares}

We assume a standard, spherical arrangement for the material ejected by the
nova during the main 2019 outburst, characterized by internal and external
radii and with the WD at the center (cf.  Figure~\ref{fig:fig8}).  The
presence of persistent [FeX] in emission, suggests that the WD was still
burning at its surface at the time of the flares, being hot and bright and
thus exerting its photo-ionizing action through the optically thin ejecta
and contrasting their recombination from higher ionized states.

Averaging over the FWHM of the emission lines and the velocity of P-Cyg
absorptions seen at the time of maximum brightness, we may adopt 1000 km/s
as the expansion velocity of the bulk of the ejecta.  In the 10 months
elapsed since the outburst in Oct 2019, at the time of the flares, the ejecta
have expanded to a radius of 170 AU, corresponding to a travel light-time of
2.0 days to cross the diameter of the shell.  In other words, an external
observer will receive news from the receding side of the ejecta only 2.0
days after being informed about the approaching one.

We believe that a flare in NSct2019 was initiated by a sudden ejection
(spherical or at least bi-conical along an axis approximately oriented to
the line of sight) of a limited about of material from the central WD.  The
FWHM=2300~km/s broad wings visible in the H$\alpha$ flare profile of
Figure~\ref{fig:figr} trace the ejected material, which mass appears much
smaller than that ejected during the main outburst, as the $\sim$1:10 ratio
in the H$\alpha$ flux suggests.  The material is optically thick when
expelled, and remain so until after the flare peak, which marks the time
when the expanding pseudo-photosphere reaches its maximum radius.  The
expanding pseudo-photosphere formed by the optically thick material causes a
drop in the surface temperature of the central photo-ionizing source,
driving a recombination wave through the ejecta.
The effect is more pronounced at inner radii of
the ejecta where the higher electron density allows the recombination to
proceed at a faster pace.  Emission from [FeVII] and [FeX] is quenched down because their
recombination is no more contrasted by photoionization from the central
source, and a surge is observed in the emission from lower
ionization lines such as Balmer, HeI, [OI], [OII], [OIII], [NII], [ArIII]
etc.  populated by recombination from higher ionization states.

\begin{figure}
	\includegraphics[width=8.6cm]{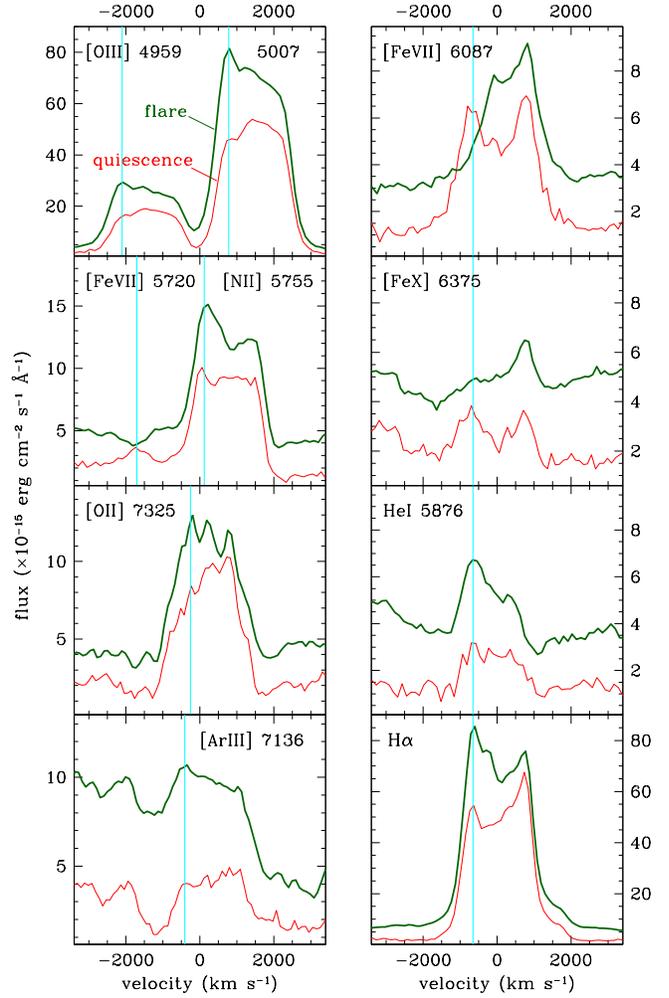}
        \caption{Comparison of the profiles of some representative emission
        lines in the spectra of Nova Sct 2019 in Figure~\ref{fig:fig6}.  The
        profiles from the flare spectrum of July 6.0 UT are plotted in green
        and thicker line (higher fluxes), those for the quiescence spectrum
        on July 5.0 UT are in red and thinner line (lower fluxes).  The
        vertical lines at $-$650~km\,s$^{-1}$ mark the position of the blue
        peak that disappear during flares for [FeVII] and [FeX] lines and
        instead reinforce in all others.}
    \label{fig:fig7}
    \end{figure}

\begin{figure}
	\includegraphics[width=\columnwidth]{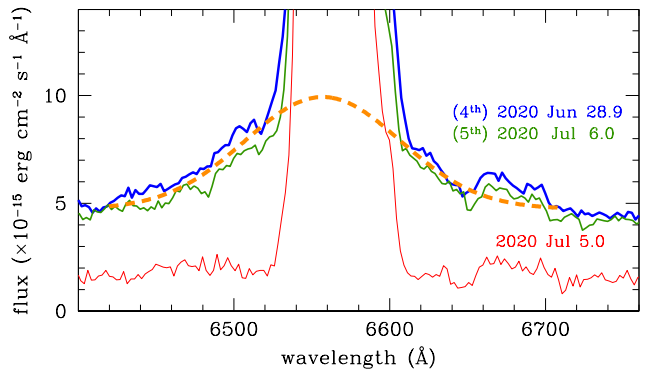}
        \caption{Zooming on the wing of H$\alpha$ for the spectra presented
        in Figure~\ref{fig:fig6}, to highlight the appearance of broad wings
        at flare peaks (Jun 28.9 and Jul 6.0) compared to the in-between quiescence
        (Jul 5.0). The dashed line is a Gaussian of FWHM=2300~km/s plotted
        for reference.}
    \label{fig:figr}
    \end{figure}

\begin{figure}
	\includegraphics[width=8.3cm]{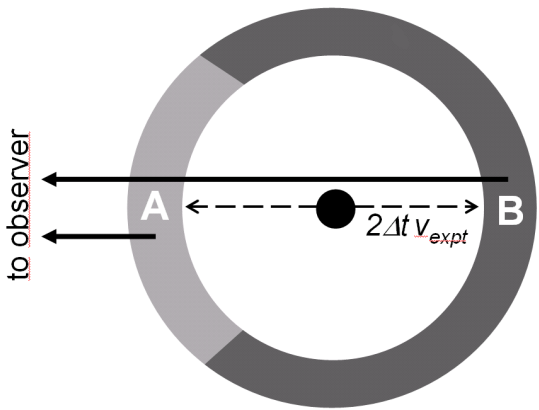}
        \caption{The simple spherical-shell geometry of the ejecta (not to
        scale) adopted in sect.~6.1 to explain the behavior during flares 
        of the profiles of emission lines.}
    \label{fig:fig8}
    \end{figure}

Our spectroscopic observations during flares (Figure~\ref{fig:fig6},
\ref{fig:fig7}, and \ref{fig:figr}) have been obtained within hours from the
recorded photometric maximum.  Within such a short time interval, only light
from the approaching ejecta has been able to reach the observer (the
light-grey $A$ portion in Figure~\ref{fig:fig8} where originates the blue
portion of the emission line profiles), while the rest of the ejecta (the
dark-grey $B$ portion in Figure~\ref{fig:fig8} that produces the rest of the
line profiles) still appear to the observer as it was {\it before} the onset
of the flare.  In the $A$ portion of the ejecta in Figure~\ref{fig:fig8},
not more exposed to hard radiation from the central WD, the recombination
depletes the medium from the highest ionization species (like [FeVII] and
[FeX]) and as a consequence the blue peak in their double-peaked profiles
fades away.  At the same time, the recombination from higher ionization
levels in the $A$ region, increases the density of lower ionization species
and boost the blue peak in their double-peaked profiles.

The return to pre-flare conditions is rather quick, the $e-$folding time for
decline in brightness after a flare peak being $\approx$50~hours (cf. 
Figure~\ref{fig:fig5}).  Unfortunately, we do not have spectra of NSct2019
obtained two or three days past the maximum brightness of a flare; we may
however predict that on such spectra the ratio of blue-to-red peaks in the
double-peaked profile would appeared reversed with respect to
Figure~\ref{fig:fig7}: the strongest peak would be the blue one for [FeVII]
and [FeX], and the red one for the other lines.

\subsection{Flaring novae}

Very few novae have presented sequences of quick flares in their outburst
lightcurve as those displayed by NSct19.

Such events should {\it not} be confused with the chaotic up-and-downs that
several novae present during their plateaued maxima, as shown by DQ~Her,
HR~Del, V723~Cas, V2540~Oph, or V1405 Cas among many others
\citep[eg.][]{1964gano.book.....P, 2004PASJ...56S.193K,
2010AJ....140...34S}, nor with the single-secondary maxima of V2362~Cyg,
V1493~Aql, or V2491~Cyg \citep[eg.][]{2004AJ....128..405V,
2008A&A...492..145M, 2009ApJ...694L.103H, 2010PASJ...62.1103A,
2011NewA...16..209M}, nor even with the (periodic) oscillation that some
novae present around the time of transition from optically-thick to -thin
conditions, like V603~Aql, V1494~Aql, or LZ~Mus
\citep[eg.][]{1960stat.book..585M, 1995cvs..book.....W, 2000NewAR..44P..65R,
2003ASPC..303..232R, 2003A&A...404..997I}.  Also the rapid variability
presented in quiescence by V2487 Oph \citep{2022MNRAS.512.1924S} or by
systems like TV~Col \citep{2022Natur.604..447S} represent entirely different
phenomena, driven by strong magnetic fields.

Our definition of a {\it flaring nova} is the following:
\begin{enumerate}
\item the flares appear superimposed on an otherwise smooth and normally
      evolving lightcurve of a nova outburst;
\item the flares are isolated and very quick events, coming in sequences;
\item a sequence of flares shows a clearly ordered pattern, either
      in terms of time-intervals and/or energy released;
\item the rise-time to flare peak brightness is rather short ($\leq$1
      day), and the exponential decline is characterized by a quick
      $e$-folding time, of the order of a few days;
\item the large amplitude of the flares ($\Delta m$$\geq$1~mag) makes them
      outstanding features of the lightcurve.
\end{enumerate}

To the best of our knowledge, there are only four novae that satisfy these
criteria: V458~Vul \citep{2015ARep...59..920T}, V4745 Sgr
\citep{2005A&A...429..599C, 2011PASJ...63..159T}, V5588~Sgr
\citep{2015MNRAS.447.1661M}, and NSct19 discussed in this paper.  All of
them started as FeII-type of slow/modest expansion velocities, with V458~Vul
and V5588~Sgr turning hybrid \citep{1992AJ....104..725W} at later times. 
All presented the [FeX] coronal emission line in their spectra, with the
exception of V4745~Sgr for which the spectroscopic monitoring may have
stopped too early to catch the high-ionization phase. Line profiles at the
time of flares suggest that a low amount of material (much smaller then 
that expelled during the initial nova eruption) is ejected at
high velocity. 

Two main characteristics put NSct19 aside from the other flaring novae:
($a$) in NSct19 the flares were observed to occur at late times, during the
advanced nebular phase, while for the other novae the flares appeared very
early in the outburst, close to maximum brightness, and ($b$) the time
interval between consecutive flares {\it decreases} in NSct19 while it {\it
increases} for the others.  There are many other differences that
contributes to make the group of flaring novae rather heterogeneous, like
that fact V458~Vul and V5588~Sgr did not develop a nebular spectrum contrary
to V4745 Sgr and NSct19, or the photometric colours during a flare evolved in
opposite directions for NSct19 and V5588~Sgr (multi-band lightcurves are not
available for V458~Vul and V4745 Sgr).

A detailed comparison of the properties of the four flaring novae is well
beyond the scopes of the present paper, but a comparative study would
certainly be instructive to carry out \citep[eg.][]{2009ApJ...701L.119P},
especially if supported by basic information like orbital periods and
inclinations, WD mass, companion type, and presence and role of magnetic
fields, all rather difficult to obtain in view of the faintness of these
novae in quiescence.

\section{Conclusions}

The observations monitoring the evolution of novae usually rarefy or
even stop when they enter and progress through the nebular stage, on the
wisdom that changes will be mostly slow, gradual, and predictable.  Our
observations of NSct19 clearly prove that this is not always the case, with
the catching of rather unexpected phenomena rewarding a persisting
observational effort.  

NSct19 displayed flaring of a nature never before detected: between days
+217 and +271 from optical maximum, nine short-lived brightenings of between
0.4 and 1.7 magnitudes in $V$ and $B$ were observed, all rather similar in
their photometric and spectroscopic development.  At the time the nova was
still burning nuclearly at the surface of the WD and well into the advanced nebular
stage, $\sim$7 mag below maximum brightness, and with optical and IR spectra
displaying forbidden lines of a high ionization degree (eg. [ArV], [FeVII],
[FeX]).  The flares appeared all of a sudden, without precursor events, and
the sequence neatly stopped after the ninth flare.  The time interval
between the flares followed an ordered sequence, declining linearly from
8.43 to 4.90 days, that safely allows to exclude that any other flare
occured without being recorded by the observations.  The color and
spectroscopic evolution of the flares indicates that their origin resides in
repeated episodes of mass ejection from the WD.

A few other novae have been noted to show flares, but they appeared very
early in the outburst, close to maximum brightness, and with the time interval
between consecutive flares increasing, while it was instead decreasing for
NSct19. Available observations not always allow to constrain the origin
of the flares, but at least for V5588~Sgr they were traced to episodic 
mass ejections from the WD, similalrly to NSct19.

\section*{Acknowledgements}

We thank the Referee (Stewart Eyres) for valuable suggestions.  We also
acknowledge the support by to P.  Valisa, P.  Ochner, and A.  Frigo to this
project.

\section{Data availability} \label{sec:data}

The data underlying this article will be shared on reasonable
request to the corresponding author.

\bsp	
\label{lastpage}
\end{document}